\documentstyle[12pt]{article}
\begin{document}
\thispagestyle{empty}
\newcommand{\be}{\begin{equation}}
\newcommand{\ee}{\end{equation}}
\newcommand{\sect}[1]{\setcounter{equation}{0}\section{#1}}
\newcommand{\vs}[1]{\rule[- #1 mm]{0mm}{#1 mm}}
\newcommand{\hs}[1]{\hspace{#1mm}}
\newcommand{\mb}[1]{\hs{5}\mbox{#1}\hs{5}}
\newcommand{\bea}{\begin{eqnarray}}
\newcommand{\eea}{\end{eqnarray}}
\newcommand{\wt}[1]{\widetilde{#1}}
\newcommand{\ux}[1]{\underline{#1}}
\newcommand{\ov}[1]{\overline{#1}}
\newcommand{\sm}[2]{\frac{\mbox{\footnotesize #1}\vs{-2}}
           {\vs{-2}\mbox{\footnotesize #2}}}
\newcommand{\prt}{\partial}
\newcommand{\eps}{\epsilon}\newcommand{\p}[1]{(\ref{#1})}
\newcommand{\R}{\mbox{\rule{0.2mm}{2.8mm}\hspace{-1.5mm} R}}
\newcommand{\Z}{Z\hspace{-2mm}Z}
\newcommand{\cd}{{\cal D}}
\newcommand{\cg}{{\cal G}}
\newcommand{\ck}{{\cal K}}
\newcommand{\cw}{{\cal W}}
\newcommand{\vj}{\vec{J}}
\newcommand{\vl}{\vec{\lambda}}
\newcommand{\vz}{\vec{\sigma}}
\newcommand{\vt}{\vec{\tau}}
\newcommand{\poiss}{\stackrel{\otimes}{,}}
\newcommand{\tx}{\theta_{12}}
\newcommand{\tb}{\overline{\theta}_{12}}
\newcommand{\zw}{{1\over z_{12}}}
\newcommand{\sqp}{{(1 + i\sqrt{3})\over 2}}
\newcommand{\sqm}{{(1 - i\sqrt{3})\over 2}}
% REVUES POUR BIBLIO
\newcommand{\NP}[1]{Nucl.\ Phys.\ {\bf #1}}
\newcommand{\PLB}[1]{Phys.\ Lett.\ {B \bf #1}}
\newcommand{\PLA}[1]{Phys.\ Lett.\ {A \bf #1}}
\newcommand{\NC}[1]{Nuovo Cimento {\bf #1}}
\newcommand{\CMP}[1]{Commun.\ Math.\ Phys.\ {\bf #1}}
\newcommand{\PR}[1]{Phys.\ Rev.\ {\bf #1}}
\newcommand{\PRL}[1]{Phys.\ Rev.\ Lett.\ {\bf #1}}
\newcommand{\MPL}[1]{Mod.\ Phys.\ Lett.\ {\bf #1}}
\newcommand{\BLMS}[1]{Bull.\ London Math.\ Soc.\ {\bf #1}}
\newcommand{\IJMP}[1]{Int.\ J.\ Mod.\ Phys.\ {\bf #1}}
\newcommand{\JMP}[1]{Jour.\ Math.\ Phys.\ {\bf #1}}
\newcommand{\LMP}[1]{Lett.\ Math.\ Phys.\ {\bf #1}}
%\renewcommand{\thefootnote}{\fnsymbol{footnote}}
%\footnotemark
\newpage
\setcounter{page}{0} \pagestyle{empty} \vs{12}
%{\large{\em Draft version}}
\begin{center}
{\LARGE {\bf Octonionic Realizations of $1$-dimensional}}\\
{\quad}\\ {\LARGE{\bf Extended Supersymmetries. A
Classification.}}\\
 [0.8cm]

\vs{10} {\large H.L. Carrion, M. Rojas and F. Toppan} ~\\ \quad
\\
 {\large{\em CBPF - CCP}}\\{\em Rua Dr. Xavier Sigaud
150, cep 22290-180 Rio de Janeiro (RJ)}\\{\em Brazil}\\

\end{center}
{\quad}\\
\centerline{ {\bf Abstract}}

\vs{6}

The classification of the octonionic realizations of the
one-dimensional extended supersymmetries is here furnished. These
are non-associative realizations which, albeit inequivalent, are
put in correspondence with a subclass of the already classified
associative representations for $1D$ extended supersymmetries.
Examples of dynamical systems invariant under octonionic
realizations of the extended supersymmetries are given. We cite
among the others the octonionic spinning particles, the $N=8$ KdV,
etc. Possible applications to supersymmetric systems arising from
dimensional reduction of the octonionic superstring and M-theory
are mentioned. \vs{6} \vfill \rightline{CBPF-NF-044/02} {\em
E-mails:}{ hleny@cbpf.br, mrojas@cbpf.br, toppan@cbpf.br}
%\newpage
\pagestyle{plain}
\renewcommand{\thefootnote}{\arabic{footnote}}
%\setcounter{footnote}{0}
%\vs{8}

\section{Introduction.}

In this work we extend the results of \cite{pt} and present a
classification of the non-associative realizations of the
$1$-dimensional extended supersymmetries which are based on the
division algebra of the octonions.\par In \cite{pt} it was proven
that the linear multiplets of representations of $1D$-Extended
Supersymmetry (i.e. the basic ingredients in Supersymmetric
Quantum Mechanics) fall into classes of equivalence each one
characterized by a given ``short multiplet", denoted as $\{n,n\}$,
such that all the $n$ bosonic (and all the $n$ fermionic) states
are grouped together with the same spin, the spin of the fermions
differing by ${\textstyle\frac{1}{2}}$ w.r.t. the spin of the
bosons. This first result of \cite{pt} was obtained by noticing
that the supersymmetry transformations of the higher-spin (let's
say at the level $s$) components in a given multiplet (these
transformations in any $D$ dimension are total derivatives) in
$D=1$ coincide with time derivatives. In its turn this implies
that the original supersymmetry multiplet can be algebraically
replaced by a shorter one, with the original higher spin states
now accommodated together with the $s-1$ spin states in this
level. The iteration of such a procedure produces at the end a
short multiplet as defined above.\par The second part of the
\cite{pt} classification consisted in proving that all short
multiplets are in one-to-one correspondence with a special class
(which can be named either of ``Weyl type" or ``of supersymmetric
type") of irreducible representations of Clifford algebras. These
are the Clifford algebra representations, splitted into $2\times
2$ block matrices, which are non-vanishing only in the
antidiagonal blocks. The restriction to this class of Clifford
algebras can be intuitively understood when thinking to them as
the Clifford algebras which can be ``promoted" to be fermionic
matrices (as it would be expected for supersymmetry), realizing a
superalgebra.\par The one-to-one correspondence between Weyl-type
Clifford algebras on one side and short multiplets of $1D$
extended supersymmetries on the other side implies the following
identifications
\begin{eqnarray}
D&=&N
\end{eqnarray}
and
\begin{eqnarray}
d&=& n
\end{eqnarray}
where on the l.h.s. $D$ denotes the spacetime dimensionality
(here, for simplicity, assumed Euclidean) of the given Clifford
algebra and $2d$ the matrix size of the representation. For what
concerns the r.h.s. $N$ denotes the number of extended
supersymmetries, while $n$ is the number of bosonic (or fermionic)
states in the given short multiplet.\par On the other hand, not
all extended supersymmetries admit a matrix representation. There
are several examples, some of them discussed in the following, of
dynamical systems admitting a non-associative realization of the
extended supersymmetries. Of course the non-associativity prevents
representing the supersymmetry transformations through standard
supermatrices and therefore these supersymmetric systems are
outside the \cite{pt} classification. It is worth mentioning that
in all the known examples the non-associativity enters through the
octonionic structure constants.\par In this paper, mimicking the
approach of \cite{pt}, we extend its results relying this time
upon the classification of the octonionic realizations for
Clifford algebras. We are able to classify the octonionic-valued
extended supersymmetries carried by octonionic-valued short
multiplets.\par The first non-trivial example of a non-associative
realization of supersymmetry involves the octonionic realization
of the $N=8$ supersymmetry. A dynamical system admitting
invariance under a global octonionic $N=8$ is e.g. given by the
$N=8$ super-KdV \cite{n8kdv}, whose Poisson brackets coincide with
the Non-associative $N=8$ Superconformal Algebra introduced in
\cite{n8sca}. The superKdVs are non-linear non-relativistic
systems in $(1+1)$-dimensions. However, since the supersymmetry
transformations depend on the space coordinate alone, they are
classified in agreement with the results for the $1D$ Extended
Supersymmetries. \par Other dynamical systems admitting invariance
under non-associative realizations of the Extended Supersymmetries
involve the octonionic spinning particles, as later discussed.
\par The scheme of this paper is as follows. In the next section
we present the classification of the octonionic realizations of
the Clifford algebras. This is the necessary ingredient for
introducing in section $3$ the classification of the octonionic
realizations of the Extended Supersymmetries. In section $4$ some
examples of dynamical systems admitting octonionic realizations of
the extended supersymmetries are discussed in some detail.
Finally, in the Conclusions, we make further comments on our
results, mentioning, among other, the possible relevance of the
classification of the one-dimensional octonionic supersymmetries
in the dimensional reduction from octonionic string and M-theory.

\section{The octonionic Clifford algebras.}

The classification of the $1D$ Extended Supersymmetries is based,
as recalled in the Introduction, on the classification of Clifford
algebras \cite{{abs},{oku}}. We summarize here the main results
which will be used in following concerning real and
octonionic-valued Clifford algebras. A very convenient
presentation for them is in terms of the following algorithm,
which allows individuating a single representative for each
irreducible class of representations of Clifford's Gamma
matrices.\par Let us prove at first that a recursive construction
of $D+2$ spacetime dimensional Clifford algebras is available,
when assumed known a $D$ dimensional representation. Indeed, it is
a simple exercise to verify that if $\gamma_i$'s denotes the
$d$-dimensional Gamma matrices of a $D=p+q$ spacetime with $(p,q)$
signature (namely, providing a representation for the $C(p,q)$
Clifford algebra) then $2d$-dimensional $D+2$ Gamma matrices
(denoted as $\Gamma_j$) of a $D+2$ spacetime are produced
according to either
\begin{eqnarray}
 \Gamma_j &\equiv& \left(
\begin{array}{cc}
  0& \gamma_i \\
  \gamma_i & 0
\end{array}\right), \quad \left( \begin{array}{cc}
  0 & {\bf 1}_d \\
  -{\bf 1}_d & 0
\end{array}\right),\quad \left( \begin{array}{cc}
  {\bf 1}_d & 0\\
  0 & -{\bf 1}_d
\end{array}\right)\nonumber\\
&&\nonumber\\ (p,q)&\mapsto&
 (p+1,q+1).\label{one}
\end{eqnarray}
or
\begin{eqnarray}
 \Gamma_j &\equiv& \left(
\begin{array}{cc}
  0& \gamma_i \\
  -\gamma_i & 0
\end{array}\right), \quad \left( \begin{array}{cc}
  0 & {\bf 1}_d \\
  {\bf 1}_d & 0
\end{array}\right),\quad \left( \begin{array}{cc}
  {\bf 1}_d & 0\\
  0 & -{\bf 1}_d
\end{array}\right)\nonumber\\
&&\nonumber\\ (p,q)&\mapsto&
 (q+2,p).\label{two}
\end{eqnarray}
As an example, the two-dimensional real-valued Pauli
matrices $\tau_A$, $\tau_1$, $\tau_2$ which realize the Clifford
algebra $C(2,1)$ are obtained by applying either (\ref{one}) or
(\ref{two}) to the number $1$, i.e. the one-dimensional
realization of $C(1,0)$. We have indeed
\begin{eqnarray}
&\tau_A= \left(\begin{array}{cc}0 &1\\ -1&0  \end{array}\right),
\quad \tau_1= \left(\begin{array}{cc}0 &1\\ 1&0
\end{array}\right), \quad
\tau_2= \left(\begin{array}{cc}1 &0\\ 0&-1  \end{array}\right).
\quad &\label{Pauli}
\end{eqnarray}
All Clifford algebras are obtained by recursively applying the
algorithms (\ref{one}) and (\ref{two}) to the Clifford algebra
$C(1,0)$ ($\equiv 1$) and the Clifford algebras of the series
$C(0, 3+4m)$ ($m$ non-negative integer), which must be previously
known. This is in accordance with the scheme illustrated in the
table below.\par {~}\par {\em Table with the maximal Clifford
algebras (up to $d=256$).}
 {\tiny{
{\begin{eqnarray} &
\begin{tabular}{|cccccccccccccccccccc}
\hline
   1  &$\ast$& 2&$\ast$&  4&$\ast$& 8&$\ast$&16&$\ast$&32&$\ast$&64&
   $\ast$&128&$\ast$&256&$\ast$
      \\ \hline
  &&&&&&&&&&&&&\\
 $\underline{(1,0)}$ &$\Rightarrow$& $(2,1)$ &$\Rightarrow$&(3,2)
 &$\Rightarrow$&
  (4,3) &$\Rightarrow$&(5,4)& $\Rightarrow$ &(6,5)
&$\Rightarrow$&
  (7,6) &$\Rightarrow$&(8,7)& $\Rightarrow$ &(9,8)
  &$\Rightarrow$

  \\
  &&&&&&&&&&\\
  &&&&&&&&&&\\
   &&&&&&(1,4)&$\rightarrow $&(2,5)&$\rightarrow$&(3,6)
&$\rightarrow $&(4,7)&$\rightarrow$&(5,8)&$\rightarrow
$&(6,9)&$\rightarrow$
   \\
  &&&&&$\nearrow$&&&&&\\
  &&&&{\underline{(0,3)}}&& &&&&\\
  &&&&&$\searrow$&&&&&\\
  &&&&&&&&&&\\
  &&&&&&(5,0)&$\rightarrow
  $&(6,1)&$\rightarrow$&(7,2)&$\rightarrow$
&(8,3)&$\rightarrow $&(9,4)&$\rightarrow$&(10,5)&$\rightarrow$
  \\
   &&&&&&&&&&\\
  &&&&&&&&&&\\
   &&&&&&&&(1,8)&$\rightarrow $&(2,9)
&$\rightarrow $&(3,10)&$\rightarrow$&(4,11)&$\rightarrow
$&(5,12)&$\rightarrow$
   \\
  &&&&&&&$\nearrow$&&&\\
  &&&&&&{\underline{(0,7)}}&& &&\\
  &&&&&&&$\searrow$&&&\\
  &&&&&&&&&&\\
  &&&&&&&&(9,0)&$\rightarrow $&(10,1)
&$\rightarrow $&(11,2)&$\rightarrow$&(12,3)&$\rightarrow
$&(13,4)&$\rightarrow$
\\
&&&&&\\ &&&\\
  &&&& &&&&&&&&&&(1,12)&$\rightarrow $&(2,13)
&$\rightarrow $
   \\
  &&&&&&&&&&&&&$\nearrow$&&&\\
  &&&&&&&&&&&&{\underline{(0,11)}}&& &&\\
  &&&&&&&&&&&&&$\searrow$&&&\\
  &&&&&&&&&&&&&&\\
  &&&&&&&&&&&&&&(13,0)&$\rightarrow $&(14,1)
&$\rightarrow $\\ &&&&&\\ &&&\\
 &&&& &&&& &&&&&&&&(1,16)&$\rightarrow $
   \\
  &&&&&&&&&&&&&&&$\nearrow$&&&\\
  &&&&&&&&&&&&&&{\underline{(0,15)}}&& &&\\
  &&&&&&&&&&&&&&&$\searrow$&&&\\
  &&&&&&&&&&&&&&\\
  &&&&&&&&&&&&&&&&(17,0)&$\rightarrow $\\

\end{tabular}&\nonumber
\end{eqnarray}}}}
\begin{eqnarray}\label{bigtable}
&& \end{eqnarray} Concerning the previous table, some remarks are
in order. The columns are labeled by the matrix size ${d}$ of the
maximal Clifford algebras. Their signature is denoted by the
$(p,q)$ pairs. Furthermore, the underlined Clifford algebras in
the table are called the ``primitive maximal Clifford algebras".
The remaining maximal Clifford algebras, known as the ``maximal
descendant Clifford algebras", are obtained from the primitive
maximal Clifford algebras by iteratively applying the two
recursive algorithms (\ref{one}) and (\ref{two}). Any Clifford
algebra is said ``non-maximal" if obtained by a maximal one by
deleting a certain number of Gamma matrices. It should be noticed
that Clifford algebras in even-dimensional spacetimes are always
non-maximal.\par For what concerns the construction of the
primitive maximal Clifford algebras of the series $C(0, 3+8n)$
(also known as quaternionic series, due to its connection with
this division algebra, as we will explain later), as well as the
octonionic series $C(0,7+8n)$, the answer can be provided with the
help of the three Pauli matrices (\ref{Pauli}). We construct at
first the $4\times 4$ matrices realizing the Clifford algebra
$C(0,3)$ and the $8\times 8$ matrices realizing the Clifford
algebra $C(0,7)$. They are given, respectively, by
\begin{eqnarray}
C(0,3) &\equiv& \begin{array}{c}
  \tau_A\otimes\tau_1, \\
  \tau_A\otimes\tau_2, \\
  {\bf 1}_2\otimes \tau_A.
\end{array}
\end{eqnarray}
and
\begin{eqnarray}
C(0,7) &\equiv& \begin{array}{c}
  \tau_A\otimes\tau_1\otimes{\bf 1}_2, \\
  \tau_A\otimes\tau_2\otimes{\bf 1}_2, \\
  {\bf 1}_2\otimes \tau_A\otimes \tau_1,\\
  {\bf 1}_2\otimes \tau_A\otimes \tau_2,\\
  \tau_1\otimes{\bf 1}_2\otimes\tau_A,\\
  \tau_2\otimes{\bf 1}_2\otimes\tau_A,\\
  \tau_A\otimes\tau_A\otimes\tau_A.
\end{array}\label{c07}
\end{eqnarray}
The three matrices of $C(0,3)$ will be denoted as ${\overline
\tau}_i$, $=1,2,3$. The seven matrices of $C(0,7)$ will be denoted
as ${\tilde \tau}_i$, $i=1,2,\ldots,7$. \par In order to construct
the remaining Clifford algebras of the series we need at first to
 apply the
(\ref{one}) algorithm to $C(0,7)$ and construct the $16\times 16$
matrices realizing $C(1,8)$ (the matrix with positive signature is
denoted as $\gamma_9$, ${\gamma_9}^2 ={\bf 1}$, while the eight
matrices with negative signatures are denoted as $\gamma_j$,
$j=1,2\ldots , 8$, with ${\gamma_j}^2 =-{\bf 1}$).  We are now in
the position to explicitly construct the whole series of primitive
maximal Clifford algebras $C(0,3+8n)$, $C(0,7+8n)$ through the
formulas
\begin{eqnarray}
C(0,3+8n)&\equiv& \begin{array}{lcr} {\overline\tau}_i\otimes
\gamma_9\otimes \ldots&\ldots&\ldots\otimes\gamma_9,\\ {\bf
1}_4\otimes\gamma_j\otimes{\bf 1}_{16}\otimes\ldots & \ldots &
\ldots\otimes{\bf 1}_{16},\\
 {\bf 1}_4\otimes\gamma_9\otimes\gamma_j\otimes {\bf
1}_{16}\otimes\ldots &\ldots&\ldots\otimes{\bf 1}_{16},  \\ {\bf
1}_4\otimes\gamma_9\otimes\gamma_9\otimes\gamma_j\otimes {\bf
1}_{16}\otimes \ldots&\ldots&\ldots\otimes{\bf 1}_{16},  \\ \ldots
&\ldots&\ldots, \\ {\bf
1}_4\otimes\gamma_9\otimes\ldots&\ldots&\otimes
\gamma_9\otimes\gamma_j,
\end{array}\label{quatern}
\end{eqnarray}
and similarly
\begin{eqnarray}
C(0,7+8n)&\equiv& \begin{array}{lcr} {\tilde\tau}_i\otimes
\gamma_9\otimes \ldots&\ldots&\ldots\otimes\gamma_9,\\ {\bf
1}_8\otimes\gamma_j\otimes{\bf 1}_{16}\otimes\ldots & \ldots &
\ldots\otimes{\bf 1}_{16},\\
 {\bf 1}_8\otimes\gamma_9\otimes\gamma_j\otimes {\bf
1}_{16}\otimes\ldots &\ldots&\ldots\otimes{\bf 1}_{16},  \\ {\bf
1}_8\otimes\gamma_9\otimes\gamma_9\otimes\gamma_j\otimes {\bf
1}_{16}\otimes \ldots&\ldots&\ldots\otimes{\bf 1}_{16},  \\ \ldots
&\ldots&\ldots, \\ {\bf
1}_8\otimes\gamma_9\otimes\ldots&\ldots&\otimes
\gamma_9\otimes\gamma_j,\label{octon}
\end{array}
\end{eqnarray}
Please notice that the tensor product of the $16$-dimensional
representation is taken $n$ times. The total size of the
(\ref{quatern}) matrix representations is then $4\times 16^n$,
while the total size of (\ref{octon}) is $8\times 16^n$.
\par
The formulas given above provide quite a practical and efficient
tool to operatively construct the irreducible Clifford algebras.
\par An important subclass of Clifford
Gamma matrices is obtained by the matrices which are decomposable
in $2\times 2$ blocks and are non-vanishing only in the
anti-diagonal blocks. Such matrices can be named as (generalized)
Weyl-type matrices (they can also be regarded of ``supersymmetric
type" since they can be promoted to be fermionic matrices
associated with the representations of the extended
supersymmetries, see \cite{top}). An inspection of the previous
tables shows that sets of (generalized) Weyl matrices are found in
special signatures only. All primitive Clifford algebras are not
of (generalized) Weyl type. However, all the derived Clifford
algebras, through the two lifting algorithms, are of Weyl-type,
once deleted the $ \left(
\begin{array}{cc}
   {\bf 1}_d &0 \\
  0 & -{\bf 1}_d
\end{array}\right)$ matrix to produce a non-maximal Clifford
algebra.\par So far we have shown how to construct the irreducible
representations of Clifford algebras, and not yet elucidated their
relations with division algebras. Such a relation can be expressed
as follows. The three matrices appearing in $C(0,3)$ can also be
expressed in terms of the imaginary quaternions $\tau_i$
satisfying $\tau_i\cdot\tau_j= -\delta_{ij}
+\epsilon_{ijk}\tau_k$. As a consequence, the whole set of maximal
primitive Clifford algebras $C(0, 3+8n)$, as well as their maximal
descendants, can be represented as quaternionic-valued matrices,
acting on spinors, which have to be interpreted now as
quaternionic-valued column vectors.\par Similarly, there exists an
alternative realization for the Clifford algebra $C(0,7)$,
obtained by identifying the seven generators with the seven
imaginary octonions satisfying the algebraic relation
\begin{eqnarray}
\tau_i\cdot \tau_j &=& -\delta_{ij} + C_{ijk} \tau_{k},
\label{octonrel}
\end{eqnarray}
for $i,j,k = 1,\cdots,7$ and $C_{ijk}$ the totally antisymmetric
octonionic structure constants given by
\begin{eqnarray}
&C_{123}=C_{147}=C_{165}=C_{246}=C_{257}=C_{354}=C_{367}=1&
\end{eqnarray}
and vanishing otherwise. This octonionic realization of the
seven-dimensional Euclidean Clifford algebra will be denoted as
$C_{\bf O}(0,7)$. Due to the non-associative character of the
(\ref{octonrel}) octonionic product (the weaker condition of
alternativity is satisfied, see \cite{gk}), the octonionic
realization cannot be represented as an ordinary matrix product
and is therefore a distinct and inequivalent realization of this
Euclidean Clifford algebra with respect to the one previously
considered (\ref{c07}). Please notice that, by iteratively
applying the two lifting algorithms to $C_{\bf O}(0,7)$, we obtain
matrix realizations with octonionic-valued entries for the maximal
Clifford algebras of the series $C(m, 7+m)$ and $C(8+m, m-1)$, for
positive integral values of $m$ ($m=1,2,\ldots$). These
realizations are denoted $C_{\bf O}(m, 7+m)$ and $C_{\bf O}(8+m,
m-1)$, respectively. The dimensionality of the corresponding
octonionic-valued matrices is $2^m\times 2^m$. \par We should
point out that the construction (\ref{octon}) leading to the
primitive maximal Clifford algebras $C(0, 7+8n)$, can be carried
on with the help of the octonionic-valued realization $C_{\bf
O}(1,8)$ for the $\gamma_i$'s and $\gamma_9$ matrices. As a
consequence, octonionic realizations of $C(0,7+8n)$ and their
descendants can be produced acting on column spinors, whose
entries are tensor products of octonions. If in the r.h.s. of
(\ref{octon}) $k$ octonionic and $n-k$ real realizations are
chosen, the maximal Clifford algebras $C(m, 7+m +8n)$ and
$C(9+8n+m, m)$, for $n\geq 0$ and $m\geq 0$, are realized by
matrices with $k+1$-tensorial octonionic entries (the extra $1$
being associated to $C_{\bf O}(0,7)$) and respective size of
$2^{4n-3k+m}$ and $2^{4n-3k+m+1}$.

\section{The octonionic extended supersymmetries.}

We furnish here the classification of the $1D$ octonionic extended
supersymmetries. More precisely, we give the list of the
octonionic maximal supersymmetries supported by short multiplets
of $n$ bosonic and $n$ fermionic fields. This is based both on
\cite{pt} and the previous section results. The octonionic
extensions can be recovered from a suitable restriction of the
\cite{pt} classification formulas of the real representations of
the $1D$ extended supersymmetries. Indeed, the octonionic
realizations of the maximal Clifford algebras are obtained by
lifting the $C_{\bf O}(0, 7+8n)$ series. No octonionic counterpart
exists for the $C(1,0)$ and the $C(0, 3+8n)$ series.\par The
one-to-one correspondence of $1D$ extended supersymmetries and
Weyl type real-valued Clifford algebras \cite{pt} is obtained by
expressing the supersymmetry generators $Q_i$ satisfying the
supersymmetry algebra
\begin{eqnarray}
\{Q_i,Q_j\}&=& \eta_{ij}H,
\end{eqnarray}
for the generalized pseudo-Euclidean metric
$\eta_{ij}$\footnote{the Euclidean case was considered earlier in
\cite{gr}. Pseudo-Euclidean supersymmetries naturally appear as
dynamical invariances for systems like the spinning particles
moving in generic space-time target manifolds.} of $(p,q)$
signature, through
\begin{eqnarray}
Q_i &=& \frac{1}{\sqrt{2}}\left(\begin{array}{cc}
  0 & \sigma_i \\
  {\tilde\sigma}_i H & 0
\end{array}\right),
\end{eqnarray}
where $H$ is the hamiltonian and one can set
\begin{eqnarray}
\Gamma_i &=& \left(\begin{array}{cc}
  0 & \sigma_i \\
  {\tilde\sigma}_i & 0
\end{array}\right)\label{weyl}
\end{eqnarray}
satisfying
\begin{eqnarray}
\Gamma_i\Gamma_j+\Gamma_j\Gamma_i &=& 2\eta_{ij}. \label{clalg}
\end{eqnarray}
The octonionic realizations are recovered by setting $\sigma_i$,
${\tilde\sigma}_i$ as matrices with octonionic-valued entries,
instead of being real matrices. From the previous section we know
that Weyl type (\ref{weyl}) octonionic-valued matrices $\Gamma_i$
satisfying (\ref{clalg}) are recovered from the maximal Clifford
algebras derived from the $C_{\bf O}(0,7+8n)$ series, after
deleting the diagonal $\left( \begin{array}{cc} {\bf 1}&0\\0&-{\bf
1}\end{array} \right)$ matrix. This leaves us the two series of
octonionic maximally extended supersymmetries of $(p,q)$
signature, namely
\begin{eqnarray}
&(m, 8+8n+m)&\label{sus1}
\end{eqnarray}
and
\begin{eqnarray}
&(8+8n+m, m), &\label{sus2}
\end{eqnarray}
for integral values $n,m\geq 0$. \par In both cases the number of
octonionic bosonic, as well as fermionic, components is given by
$2^{n+m}$. Please notice the equivalence of (\ref{sus1}) and
(\ref{sus2}) under the sign flipping $p\leftrightarrow q$.\par
This result can be summarized as follows. The inequivalent classes
of octonionic irreducible realizations of the maximally extended
supersymmetries acting on octonionic multiplets of $d=2^k$ bosons
and equal number of fermions is given by
\begin{eqnarray}
(x+8\varepsilon(k+1-x), x+8(1-\varepsilon)(k+1-x) ),
\end{eqnarray}
for integral values $0\leq x\leq k$ and $\varepsilon = 0,1$.\par
At the lowest order of $d$, the following table can be produced
\begin{eqnarray}
\begin{array}{|c|c|} \hline
  & (p,q) \\ \hline
 d=1 & (8,0), (0,8)\\ \hline
 d=2& (16,0), (9,1), (1,9), (0,16) \\ \hline
 d=4 & (24,0), (17,1), (10,2), (2,10), (1,17), (0,24) \\ \hline
 d=8 & (40,0), (33,1), (26,2),(19,3), (12,4), (4,12),(3,19),
 (2,26), (1,33),(0,40)\\\hline
\end{array}
\end{eqnarray}
Of course, irreducible realizations of non-maximal octonionic
extended supersymmetries are recovered from the previous table for
the values $(p',q')$, with $p'\leq p$ and $q'\leq q$, provided
that $p'$, $q'$ are not too small. For instance, the irreducible
$2+2$ realization of the octonionic $(8,1)\subset (9,1)$ is
encountered, while the irreducible $(8,0)$ is directly present in
the table and found at $d=1$.

\section{Dynamical systems with octonionic supersymmetry.}

In this section we present some examples of dynamical systems
admitting invariance under $1D$ octonionic extended
supersymmetries.\par The first example of a non-associative,
octonionic realization of a $1D$ supersymmetry is given by the
octonionic $N=8$. It is associated, according to the previous
section results, to $C_{\bf O}(8,0)$ and is expressed in terms of
$2\times 2$ octonionic-valued matrices. Explicitly, the eight
supersymmetry generators $Q_0$, $Q_i$, for $i=1,2,\ldots,7$, are
given by
\begin{eqnarray}
Q_0 = \frac{1}{\sqrt{2}}\left(\begin{array}{cc}
  0 & 1 \\
  H & 0
\end{array}\right), &\quad &
Q_i = \frac{1}{\sqrt{2}}\left(\begin{array}{cc}
  0 & t_i \\
  -t_i H & 0
\end{array}\right),\label{superoct}
\end{eqnarray}
where $t_i$ denote the imaginary octonions and $H$ is the
hamiltonian.\par The above supersymmetry can also be expressed in
an octonionic language. It corresponds to the simplest (for $D=1$)
case of a class of higher-dimensional generalized octonionic
supersymmetries, investigated in the light of superstring
theories, M-theory, etc., see \cite{lt}. We can indeed introduce
the octonionic supercharge ${\cal Q}$ and its octonionic conjugate
${\cal Q}^\ast$ (under octonionic principal conjugation), through
\begin{eqnarray}
{\cal Q} &=& Q_0 +\frac{1}{\sqrt{7}}\sum_{i=1,\ldots 7}
Qt_i,\nonumber\\ {\cal Q}^\ast &=& Q_0
-\frac{1}{\sqrt{7}}\sum_{i=1,\ldots 7} Qt_i,\label{qoct}
\end{eqnarray}
with \begin{eqnarray} Q&=&
\frac{1}{\sqrt{2}}\left(\begin{array}{cc}
  0 & 1 \\
  -H & 0
\end{array}\right).
\end{eqnarray}
As a consequence, the octonionic $N=8$ can be rewritten as
\begin{eqnarray}
&\{{\cal Q}, {\cal Q}\}=\{{\cal Q}^{\ast},{\cal
Q}^{\ast}\}=2H,&\nonumber\\ &\{{\cal Q}, {\cal Q}^{\ast}\} = 0.&
\end{eqnarray}
We already pointed out that the octonionic $N=8$ is an
inequivalent realization of the $1D$ $N=8$ supersymmetry with
respect to standard $N=8$, obtained by replacing the seven
imaginary octonions $t_i$ in (\ref{superoct}) with the seven
(associative) $8\times 8$ matrices given in
(\ref{c07})\footnote{In the $Q_0$, $Q_i$ entries $1$ is also
replaced by ${\bf 1}_8$.}.\par  Perhaps the most convenient way of
getting ourselves convinced of the inequivalence of the
associative-versus-nonassociative realizations of $N=8$, consists
in presenting a dynamical system which only admits invariance
under the octonionic $N=8$. No counterpart is found with
invariance under the associative $N=8$. A nice example of that is
given by the $N=8$ KdV. Due to the absence of central extension
for $N$-extended superconformal algebras with $N>4$ \cite{gls},
superKdV equations only exist for $N\leq 4$. Indeed, the Virasoro
central extension is necessary to produce the three-derivative
term entering the KdV equation. On the other hand, the
mathematical no-go theorem preventing the construction of superKdV
equations for $N>4$ can be overcome by noticing that non-Jacobian
superconformal algebras like the Non-associative $N=8$ SCA
introduced in \cite{n8sca}, can present central extension and be
regarded as generalized Poisson brackets for a non-associative
supersymmetric extension of KdV. In \cite{n8kdv} we proved that
there exists only one such extension, the $N=8$ KdV, invariant
under the global octonionic $N=8$. From the considerations above,
it is clear that no $N=8$ superKdV based on the associative $N=8$
can exist, since this is prevented by the no-go theorem.\par The
$N=8$ superKdV equations are explicitly given by
\begin{eqnarray}
{\dot T} &=&- T''' - 12 T' T - 6 Q_a'' Q_a +4 J_i''J_i,\nonumber\\
{\dot Q} &=& -Q''' -6 T' Q -6 T Q' - 4 Q_i''J_i +2 Q_i
J_i''-2Q_i'J_i',\nonumber\\ {\dot Q}_i &=& -{Q_i}''' - 2 Q J_i'' -
6 TQ_i' - 6 T' Q_i + 2 Q' J_i' +4 Q''J_i - \nonumber\\ && 2
C_{ijk} ( Q_j J_k'' -Q_j' J_k' -2 Q_j''J_k ),\nonumber\\ {\dot
J}_i &=& -{J_i}''' - 4 T'J_i - 4 T J_i' + 2 Q Q_i '+2Q'Q_i
-C_{ijk}( 4 J_j J_k'' +2Q_j Q_k'). \label{eomn8}
\end{eqnarray}
(the dot and the prime denote, as usual, the time and respectively
the space derivative). They involve the eight bosonic fields $T$,
$J_i$ and the eight fermionic fields $Q_0\equiv Q$, $Q_i$
($i=1,\ldots ,7$, while $a=0,1,\ldots , 7$). One should notice the
presence of the octonionic structure constants $C_{ijk}$. The
$N=8$ global supersymmetry transformations leaving the
(\ref{eomn8}) system invariant are generated by $\int dx Q_a(x)$,
for $a=0,1,\ldots, 7$, under the Non-associative $N=8$ SCA Poisson
brackets, see \cite{n8kdv} for details. They coincide with the
(\ref{superoct}) transformations once setting the hamiltonian
$H=i\frac{\partial}{\partial x}$. Please notice that, despite the
fact that the fields entering the $N=8$ superKdV are dependent on
both the space and time coordinates, the $N=8$ supersymmetry
transformations only depend on the space coordinate $x$. The time
dependence being ``frozen", one can directly read these
transformations from the $1D$ octonionic $N=8$ supersymmetry
generators given above.\par We have clarified the role of the
octonionic supersymmetry transformations and presented a first
example of a non-trivial system invariant under non-associative
supersymmetries. Another example of a class of dynamical systems
invariant under octonionic supersymmetry involves the octonionic
spinning particles, whose simplest example is again found for
$N=8$. The octonionic generalization of the free real-valued
spinning particles is described by the octonionic-valued bosonic
and fermionic fields, $x=x_0+\sum_ix_i\tau_i$ and $\psi
=\psi_0+\psi_i\tau_i$ respectively. The imaginary octonionic
super-coordinates $x_i$, $\psi_i$ can be regarded as
super-coordinates associated with the seven sphere $S^7$ \cite{cp}
since the latter can be described by unitary octonions. The free
kinetic action is given by \begin{eqnarray} S &=& \frac{1}{2}\int
dt \cdot tr\{ (x^\ast,\psi^\ast) \left(\begin{array}{cc}
  \frac{d^2}{dt^2}& 0 \\
  0& \frac{d}{dt}
\end{array}\right) \left(\begin{array}{c}x\\\psi
\end{array}\right)\},
\end{eqnarray}
where $tr$ denotes both the matrix trace $Tr$ and the projection
over the octonionic identity \cite{n8aff}, while ``$\ast$" denotes
the octonionic principal conjugation (see also formula
(\ref{qoct})).\par The above free action is invariant under the
octonionic $N=8$ supersymmetry, whose suitably normalized
explicitly transformations acting on the component fields are
given by
\begin{eqnarray}&&
\begin{array}{ll}
  \delta_0 x_0=\psi_0, & \delta_0x_i =\psi_i, \\
  \delta_0\psi_0={\dot x}_0, & \delta_0\psi_i={\dot x}_i
\end{array}
\end{eqnarray}
and
\begin{eqnarray}&&
\begin{array}{ll}
  \delta_i x_0=-\psi_i, & \delta_ix_j =\delta_{ij}\psi_0-C_{ijk}\psi_k, \\
  \delta_i\psi_0={\dot x}_i, & \delta_i\psi_j=-\delta_{ij}{\dot
  x}_0+C_{ijk}{\dot x}_k,
\end{array}
\end{eqnarray}
with $i=1,\ldots, 7$.\par The classification of the octonionic
spinning particles, invariant under generalized $(p,q)$
supersymmetries, is an immediate consequence of the classification
formulas for the octonionic supersymmetries presented in the
previous section. The construction of the octonionic spinning
particles straightforwardly follows the one here presented for the
$N=8$, i.e. the ($p=8,q=0$), octonionic supersymmetry.

\section{Conclusions.}

In this work we furnished the classification of the octonionic
$1D$ extended supersymmetries acting on small multiplets of $n$
bosonic and $n$ fermionic fields.  The key observation allowing us
to classify the octonionic supersymmetries consists in noticing
that they are in one-to-one correspondence with the class of Weyl
type realizations of Clifford algebras, expressed through matrices
with octonionic-valued entries. ``Weyl type" simply means here the
subclass of matrices in a Clifford algebra which can be
``promoted" to be fermionic (i.e. odd-graded) elements in a
superalgebra.\par The classification of the octonionic
supersymmetries can therefore be extracted from the classification
of the octonionic Clifford algebras. Explicit tables, expressing
the number of generalized $(p,q)$ supersymmetries (for $p$
positive and $q$ negative eigenvalues) supported by the $n+n$
field multiplets, are given.
\par
We further mentioned that the octonionic realizations can be put
in correspondence with a subclass of the associative
representations of the $1D$ extended supersymmetries. Basically,
this can be done by replacing the seven imaginary octonions with
the seven antisymmetric matrices producing the Euclidean Clifford
algebra $C(0,7)$. Nonassociative and associative realizations of
supersymmetry remain, nevertheless, inequivalent. This point is
better understood by noticing that $N=8$-octonionic invariant
systems, such as the $N=8$ KdV, indeed exist, while on the other
hand is known (due to a mathematical no-go theorem discussed in
the previous section) that no $N=8$ extension of KdV based on the
associative $N=8$ supersymmetry can be constructed.\par We further
explicitly discussed another example of a class of dynamical
systems invariant under the octonionic supersymmetry, i.e. the one
provided by the octonionic spinning particles.\par Finally, it is
worth mentioning the possible relevance in physical applications
of the supersymmetric systems investigated in this paper. While it
is well known since the work of \cite{kt} that division algebras
are associated with extended supersymmetries, quite understandably
due to the complications arising from the non-associativity,
octonionic realizations received less attention than the
associative division algebras. Nevertheless, octonions continued
being investigated as, e.g., in \cite{{fm},{cs}} in the context of
superstring theory. More recently, in \cite{lt}, the existence of
an octonionic version of the M-theory with surprising features,
among the others the equivalence of the $M5$ five-brane sectors
with the $M1$ and $M2$ sectors, was pointed out. In general
higher-dimensional octonionic theories have peculiar features, for
instance they are no longer invariant under the full Lorentz
group, but under its $G_2$ coset, since this is the group of the
octonionic automorphisms.\par For what concerns the specialization
to $D=1$, i.e. the case treated in this paper, we should mention
that in the Jordan framework and at least for the restricted class
of Jordan algebras, see \cite{grp}, a consistent octonionic
quantum mechanics is available. On the other hand it is clear that
higher-dimensional octonionic supersymmetric theories, like the
superstring or the $M$-theory mentioned above, can be
dimensionally reduced to $1D$. In this passage we obtain
octonionic quantum mechanical systems admitting, as in the
standard associative case \cite{restr}, extended number of
supersymmetries. Such systems must be constructed in terms of the
octonionic multiplets here classified, leaving room for promising
applications of the results and the techniques presented in this
paper.

\end{document}